\documentclass[11pt]{article}
\usepackage[dvips]{graphicx}
\usepackage{amsmath}

\author{M. G\'omez-Bock\\
{\small \it Instituto de F\'{\i}sica, Benem\'erita Universidad
Aut\'onoma de Puebla},\\
{\small \it Apartado Postal J-48 Puebla, Pue. M\'{e}xico}}

\title{Deviations from universality of slepton masses in the MSSM}
\begin{document}
\maketitle

\begin{abstract}
In this paper we propose an ansatz that applies to the slepton mass
matrices. In our approach these matrices contain a dominant sector
that can be diagonalized exactly. We study the numerical results for
the slepton mass eigenstates, looking for deviations from
universality, which is usually assumed when one evaluates the
production of sleptons at future colliders.
\end{abstract}

\section{Introduction}
Although the MSSM is the leading candidate for new physics beyond
the Standard Model and sensible explains electroweak symmetry
breaking by stabilizing the energy scale, it still leaves with no
answer the open problems of the SM, among them the flavor
problem \cite{FN}\cite{Diaz-Cruz:2001gf}. Furthermore, SUSY brings a new flavor
problem which is closely related to the mass generation mechanism of the
superpartners. Namely, a generic sfermion mass could lead
to unacceptable large FCNC, which would exclude the
model\cite{glashow,FCNC,segundo}. Several conditions or scenarios
have been proposed to solve this problem, which reduce the number
of free parameters and safely fit the experimental restrictions. The solutions handle in the
literature include\cite{Diaz-Cruz:2005ri}:\\
\begin{itemize}
\item {\it i)} {\em
degeneration}, where different sfermion families have the same mass;

\item {\it ii)} {\em proportionality}, here the trilinear $A-terms$ are
proportional to the Yukawa couplings (SUGRA)\cite{Hall}; 

\item {\it iii)} {\em decoupling}, where the superpartners are too heavy to affect
the low energy physics(Split SUSY, focus point SUSY, inverted
hierarchy)\cite{Feng}; 

\item {\it iv)} {\em alignment}, in this case the
same physics that explains the fermion mass spectra
and mixing angles also would explain the pattern of sfermions mass spectra\cite{Nir:1993mx}.\\
\end{itemize}

In the MSSM the particle mass spectra depends on the SUSY breaking
mechanism. The parametrization of SUSY breaking for MSSM is called
{\it Soft SUSY Breaking, SSusyB}. The scalar fields are grouped in a
supermultiplet together with the fermions fields, in such a way that
the scalar masses are linked to the SSusyB energy scale\footnote{see
for instance \cite{Weinberg}} and the mass degeneracy
could be broken by the SSusyB mechanism.\\

In this paper we are going to study the slepton mass matrices. Our
goal is to determine the slepton mass eigenvalues, which are the
ones that hopefully will be measured at coming (LHC) and future
colliders (ILC). For this, we shall propose a hierarchy within the
mass matrices, which will include a sector that will have the
property of being exactly diagonalizable. This sector will determine
mostly the slepton masses. We also include a sector with small
off-diagonal entries that will lead to lepton flavor violation
(LFV), but we leave this last analysis to future work.\\

The organization of this papers is as follows. In  the next
section we present the terms that contribute to the slepton mass
matrix in the MSSM. Section 3 explicitly shows the
ansatz proposed for the trilinear terms that contribute to this mass
matrix as two contribution orders mentioned above, obtaining the
expressions for the slepton masses. We present the numerical results
for the parameter space in section 4. And finally, in section 5 we
summarize our conclusions.\\

\section{Slepton Mass Matrix}

The SUSY invariant terms, which contribute to the diagonal elements
of the mass matrix, come from the auxiliary fields, namely the $F$
and $D-terms$. However, the mass matrix also includes terms that come
from the Soft SUSY Lagrangian \cite{Martin,Kuroda:1999ks}. Within
the MSSM, this soft Lagrangian includes the following terms
\begin{equation}
\mathcal{L}_{soft}= \mathcal{L}_{sfermion}^{mass}+
\mathcal{L}_{bino}^{mass}+\mathcal{L}_{gaugino}^{mass}+\mathcal{L}_{gluino}^{mass}+\mathcal{L}_{Higgsino}^{mass}+
\mathcal{L}_{H\tilde{f}_{i}\tilde{f}_{j}} \label{softlag}
\end{equation}

In order to establish the free parameters of the model coming from
this Lagrangian, we write down the form of the slepton masses and
the Higgs-slepton-slepton couplings, the first and last term of eq.
(\ref{softlag}), which are given as
\begin{equation}
\mathcal{L}_{soft}^{\tilde{l}}=-m_{\tilde{E},ij}^{2}\tilde{\bar{E}}^{i}\tilde{\bar{E}}^{j\dag}
-m_{\tilde{L},ij}^{2}\tilde{L}^{i\dag}\tilde{L}^{j}-(A_{e,ij}\tilde{\bar{E}}^{i}\tilde{L}^{j}H_{1}+h.c)
\end{equation}

\noindent where the \textit{trilinear terms}, or \textit{A-terms},
are the
coefficient of the scalar Higgs-sfermions couplings.\\

In principle, any scalar with the same quantum numbers could mix
through the soft SUSY parameters\cite{Aitchison}. This general
mixing includes the parity superpartners fermionic labels, and leads
us to a sfermion mass matrix given as a squared $6\times 6$
matrix, which can be written as a block matrix as\\

\bigskip
\begin{equation}
\tilde{M}_{\tilde{f}}^{2}=
\begin{pmatrix}
    M_{LL}^{2} & M_{LR}^{2} \\
    M_{LR}^{2\dag} & M_{RR}^{2}
\end{pmatrix}
\label{massLR}
\end{equation}
\noindent where
\begin{eqnarray}
 M_{LL}^{2}& =&
m_{\tilde{L}}^{2}+M_{l}^{(0)2}+\frac{1}{2}\cos2\beta(2m_{W}^{2}-m_{Z}^{2})
\text{{\bf I}}_{3\times 3},\\
M_{RR}^{2} & = &
M_{\tilde{E}}^{2}+M_{l}^{(0)2}-\cos2\beta\sin^{2}\theta_{W}m_{Z}^{2}
\text{{\bf I}}_{3\times 3},\\
M_{LR}^{2} & = &
\frac{A_{l}v\cos\beta}{\sqrt{2}}-M_{l}^{(0)}\mu\tan\beta.
\label{Aterm}
\end{eqnarray}
\noindent where $M_{l}^{(0)}$ is the lepton mass matrix.\\

The lepton-flavor conservation is violated by the non-vanishing
off-diagonal elements of each matrix, and the size of such
elements is strongly constrained from experiments. In the SUSY
Standard Model based on supergravity, it is assumed that the mass
matrices $m_{\tilde{E}}^{2}$ and $m_{\tilde{L}}^{2}$ are
proportional to unit matrix, while $A_{e,ij}$ is proportional to the
Yukawa matrix $y_{e,ij}$. With these soft terms, the lepton-flavor
number is conserved exactly\cite{Hisano:1995nq}. However, in general
soft-breaking schemes, we expect that some degree of flavor violation
would be generated. A particular proposal for this pattern is presented next.\\

\section{An ansatz for the mass matrix}

The trilinear terms come directly from the Soft SUSY breaking terms,
and contribute toward to increase the superparticles masses. We analyze the
consequences on sfermion masses by assuming that such terms would
acquire an specific flavor structure, which is represented by some
{\it textures}. Textures represent an {\it a priori}
assumption\cite{Xing},\cite{fourtext}, in this case, for the
mixtures between sfermion families. Such a structure implies that we
can classify the matrix elements into three groups, the ones that
contribute at leading order, those that could generate
appreciable corrections and those that could be discarded,
obtaining a
hierarchal textures form. \\

We propose an ansatz for the trilinear A-terms in the flavor
basis, and study its effects on the physical states. We work on a
scheme that performs exact diagonalization. First, we parameterize
off-diagonal terms assuming a favor asymmetry inherited from the
fermionic SM sector. In general, there is no reason to expect that
the sfermion mass states are exactly degenerate, and there is no
solid theoretical basis to consider such pattern,
although they are phenomenologically viable\cite{Weinberg,Haber}. \\

We assume, as in supergravity models, the condition of degeneracy on
pure Left and pure Right contributions:
\begin{equation}
 M_{LL}^{2} \simeq M_{RR}^{2} \simeq \tilde{m}_{0}^{2}
\mathbf{I}_{3\times 3},
\end{equation}

Our ansatz for the $A-terms$ is build up using textures forms and
hierarchal structure as we pointed above. The parametrization is obtained by assuming
that the mixing between third and second families is larger than the
mixing with the first family. Furthermore, 
current data mainly suppress the FCNCs associated with the first two
slepton families, but allow considerable mixing between the second and third slepton families\cite{Diaz-Cruz:2001gf}. \\

Thus, our proposal includes dominant terms that mix the second and
third families, as follows

\begin{equation}
A_{LO}= A'_{l}=
\begin{pmatrix}
  0 & 0 & 0 \\
  0 & w & z \\
  0 & y & 1
\end{pmatrix}
A_{0}, \label{BLO}
\end{equation}

\noindent then mixtures with the first family are treated as corrections, and are given as:\\

\begin{equation}
\delta A_{l}=
\begin{pmatrix}
 e & s & r \\
s & 0 & 0 \\
  r & 0 & 0
\end{pmatrix}
A_{0}=
\begin{pmatrix}
 \delta A_{e} & \delta A_{s} & \delta A_{r} \\
\delta A_{s} & 0 & 0 \\
  \delta A_{r} & 0 & 0
\end{pmatrix}
\label{BNLO}
\end{equation}\\

In the case of $w = 0$ we reproduce the ansatz given in
Ref. \cite{Diaz-Cruz:2001gf}. The dominant terms give a $4\times 4$
decoupled block mass matrix, in the basis
${\tilde{e}_{L},\tilde{e}_{R},
\tilde{\mu}_{L},\tilde{\mu}_{R},\tilde{\tau}_{L},\tilde{\tau}_{R},}$
as\\

\begin{equation}
\tilde{M}_{\tilde{l}}^{2}=\left(
\begin{array}{cc|cccc}
  a & 0 & 0 & 0 & 0 & 0 \\
  0 & a & 0 & 0 & 0 & 0 \\
  \hline
  0 & 0 & a & X_{2} & 0  & A_{z} \\
  0 & 0 & X_{2} & a & A_{y} & 0 \\
  0 & 0 & 0 & A_{y} & a & X_{3} \\
  0 & 0 & A_{z} & 0 & X_{3} & a
\end{array}\right),
\label{mBLObloq}
\end{equation}
\noindent with $X_{3}=\frac{1}{\sqrt{2}}A_{0}v\cos\beta-\mu
m_{\tau}\tan\beta$ and  $X_{2}=A_{w}-\mu m_{\mu}\tan\beta$. Where
$\mu$ is the $SU(2)-invariant$ coupling of two different Higgs
superfield doublets, $A_{0}$ is the trilinear coupling scale and
$\tan\beta=\frac{v_{2}}{v_{1}}$ is the ratio of the two vacuum
expectation values
coming from the two neutral Higgs fields, these three are MSSM parameters\cite{Aitchison,Accomando}. \\

The correction takes the form:\\
\begin{equation}
\delta \tilde{M}_{\tilde{l}}^{2}=\left(
\begin{array}{cc|cccc}
 0 & \delta A_{e} & 0 & \delta A_{s} & 0 & \delta A_{r} \\
 \delta A_e & 0 & \delta A_{s} & 0 & \delta A_{r} & 0 \\
 \hline
 0 & \delta A_{s} & 0 & 0 & 0 & 0 \\
 \delta A_{s} & 0 & 0 & 0 & 0 & 0 \\
 0 & \delta A_{r} & 0 & 0 & 0 & 0 \\
 \delta A_{r} & 0 & 0 & 0 & 0 & 0
\end{array}
\right) \label{mBNLObloq}
\end{equation}\\
The explicit forms of $A_{z,y,w}$ and $\delta A$ are given in table
\ref{formadeBi}.\\

\begin{table}[hbt]
\renewcommand{\arraystretch}{1.5}
\begin{center}
\begin{tabular}{|c|c|}
\hline
dominant & correction  \\
\hline
 $A_{z}=\frac{1}{\sqrt{2}}z A_{0}v\cos\beta$ & $\delta
A_{s}=\frac{1}{\sqrt{2}}sA_{0}v\cos\beta$
 \\
 $A_{y}=\frac{1}{\sqrt{2}}y A_{0}v\cos\beta$
& $\delta
A_{r}=\frac{1}{\sqrt{2}} rA_{0}v\cos\beta$\\
$A_{w}=\frac{1}{\sqrt{2}}w A_{0}v\cos\beta$
& $\delta A_{e}=0$\\
 \hline
\end{tabular}
\renewcommand{\arraystretch}{1.0}
\caption[]{\label{formadeBi} \it Explicit terms of the sfermion mass
matrix ansatz, assuming $\delta A_{e}$ as a third order element.}
\end{center}
\end{table}
In order to obtain the physical slepton eigenstates, we diagonalize
the $4 \times 4$ mass sub-matrix given in (\ref{mBLObloq}).For
simplicity we consider that $z=y$, which
represent that the mixtures $\tilde{\mu}_L \tilde{\tau}_R$ and
$\tilde{\mu}_R \tilde{\tau}_L$
are of the same order . The
rotation will be performed to this part using an hermitian matrix
$Z_{l}$, such that

\begin{equation}
Z_{l}^{\dag}M_{\tilde{l}}^{2}Z_{l}= \tilde{M}^{2}_{Diag},
\label{Mdiag}
\end{equation}
\noindent where
\begin{equation}
M_{\tilde{l}}^{2}=\left(
\begin{array}{cccc}
 \tilde{m}_{0}^{2} & X_2 & 0 & A_y \\
 X_2 & \tilde{m}_{0}^{2} & A_y & 0 \\
 0 & A_y & \tilde{m}_{0}^{2} & X_3 \\
 A_y & 0 & X_3 & \tilde{m}_{0}^{2}
\end{array}
\right).
\end{equation}\\

Then the rotation matrix is given by

\begin{equation}
\begin{pmatrix}
  \tilde{e}_{L}\\
  \tilde{\mu}_{L} \\
  \tilde{\tau}_{L}\\
  \tilde{e}_{R}\\
  \tilde{\mu}_{R}\\
  \tilde{\tau}_{R}
\end{pmatrix}
=\frac{1}{\sqrt{2}}\left(
\begin{array}{cccccc}
 1 & 0 & 0 & 0 & 0 & 0\\
 0 & -\sin \frac{\varphi}{2} & -\cos \frac{\varphi}{2} & 0 & \sin \frac{\varphi}{2} & \cos \frac{\varphi}{2}\\
 0 & \cos \frac{\varphi}{2} & -\sin \frac{\varphi}{2} & 0 & -\cos \frac{\varphi}{2} & \sin \frac{\varphi}{2}\\
 0 & 0 & 0 & 1 & 0 & 0\\
 0 & -\sin \frac{\varphi}{2} & \cos \frac{\varphi}{2} & 0 & -\sin \frac{\varphi}{2} & \cos \frac{\varphi}{2}\\
 0 & \cos \frac{\varphi}{2} & \sin \frac{\varphi}{2}& 0 & \cos \frac{\varphi}{2} & \sin \frac{\varphi}{2}\\
\end{array}
\right)
\begin{pmatrix}
  \tilde{e}_{1}\\
  \tilde{\mu}_{1}\\
  \tilde{\tau}_{1}\\
  \tilde{e}_{2}\\
  \tilde{\mu}_{2}\\
  \tilde{\tau}_{2}
\end{pmatrix}
= Z_{B}^{l}\tilde{l}, \label{rotationB}
\end{equation}\\

\noindent with
\begin{eqnarray}
sin\varphi & =& \frac{2A_y}{\sqrt{4 A_y^2+\left(X_2 -X_3 \right)^2}},\nonumber \\
& & \nonumber \\
cos \varphi & =&  \frac{\left(X_2 -X_3 \right)}{\sqrt{4
A_y^2+\left(X_2-X_3 \right)^2}} \label{fi}
\end{eqnarray}

\noindent  We obtain the following hierarchy for the sleptons
$m_{\tilde{\tau}_{1}}< m_{\tilde{\mu}_{1}} < m_{\tilde{\mu}_{2}} <
m_{\tilde{\tau}_{2}}$, for $\mu <0$. Having the following
eigenvalues

\begin{eqnarray}
m^{2}_{\tilde{\mu_{1}}}& = & \frac{1}{2}(2 \tilde{m}_{0}^{2}+X_{2}+X_{3}-R),\nonumber\\
m^{2}_{\tilde{\mu_{2}}}& = & \frac{1}{2}(2 \tilde{m}_{0}^{2}-X_{2}-X_{3}+R),\nonumber\\
m^{2}_{\tilde{\tau_{1}}}& = & \frac{1}{2}(2 \tilde{m}_{0}^{2}-X_{2}-X_{3}-R),\nonumber\\
m^{2}_{\tilde{\tau_{2}}}&=&\frac{1}{2}(2\tilde{m}_{0}^{2}+X_{2}+X_{3}+R),
\label{BLOmasses}
\end{eqnarray}

\noindent with $R=\sqrt{4 A_y^2+\left(X_{2}-X_3 \right)^2}$.\\

\section{Numerical results for slepton masses}

\begin{figure}[hbt!]
\begin{center}
\renewcommand{\arraystretch}{0.3}
\framebox {\includegraphics[width=13cm,height=12cm]{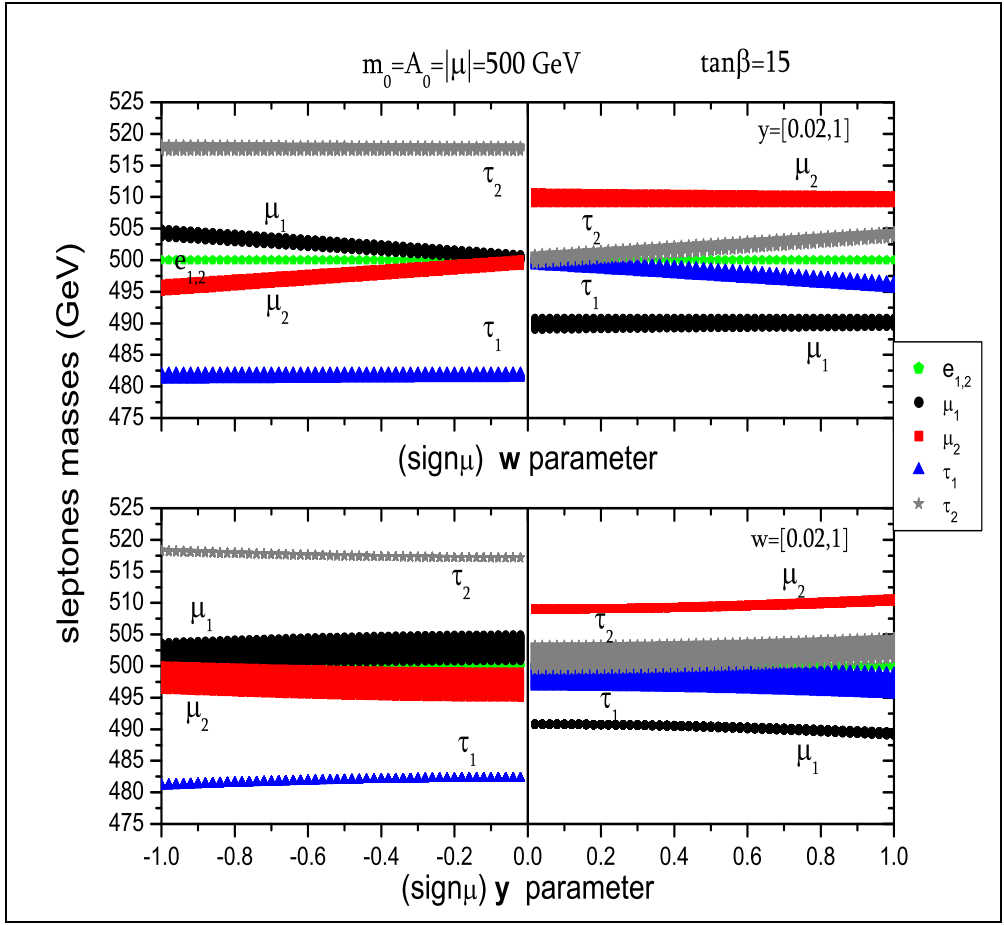}}
\caption[]{\label{fg:Bwy} {\it {\small Slepton masses dependency
with respect to the parameter ansatz $w$ (up) and $y$ (down) with
$\tilde{m}_{0}=A_{0}=\mu_{susy}=500 GeV$ and $\tan\beta=15$,
considering $\mu_{susy}<0$ and $\mu_{susy}>0$.}}}
\end{center}
\renewcommand{\arraystretch}{0.5}
\end{figure}

From the expressions for the slepton masses (eq. \ref{BLOmasses}),
we shall analyze their parameter dependency. In figure \ref{fg:Bwy}
we show the dependence on $y(=x)$ and $w$. Then, in the next two
figures we show the dependence of the slepton masses on the usual
MSSM parameters, $\mu$, $A_{0}$ and $\tan\beta$. \\

We see that $X_{3}$ and $X_{2}$ are given in terms of $\mu$ and
$\tan \beta$, having a strong dependency on the sign of $\mu$, and
so we obtain a hierarchy of the slepton masses given as follows:
\begin{eqnarray}
\mu < 0 &  & m_{\tau_{1}}<m_{\mu_{2}}<(m_{e_{1}}=m_{e_{2}})<m_{\mu_{1}}<m_{\tau_{2}}\\
\mu > 0 & &
m_{\mu_{1}}<m_{\tau_{1}}<(m_{e_{1}}=m_{e_{2}})<m_{\tau_{2}}<m_{\mu_{2}}
\end{eqnarray}

We observed this on the graphs of figure \ref{fg:Bwy}, where we run
independently the values of $y$ and $w$ in a range of $[0.02,1]$ and
set the values for the soft susy breaking scale as
$\tilde{m}_{0}=500\, GeV$, with $\tan \beta =15$. We have
practically no dependence on parameter $y$. For $w=0$ we have
degeneracy of the four lightest sleptons, and practicaly no dependency
on these parameters for the
heaviest two sleptons.
The non-degeneracy increases up to $10\,GeV$ with $w=\pm 1$ for the two middle sleptons
smuons (or staus). This result tells us that if we are to explore the
mixtures on the second and third families we have to take into
account the term coming from the smuon mass term, represented here
whit paremeter $w$, (\ref{BLO}).
As we said the strongest dependency comes from the MSSM parameter, and the
deviation from universality is manifested by the staus, which in the
case of $\mu <0$, show a difference in staus masses of $\sim 40 \,GeV$.\\

\begin{figure}[hbt!]
\begin{center}
\framebox {\includegraphics[width=12cm,height=10cm]{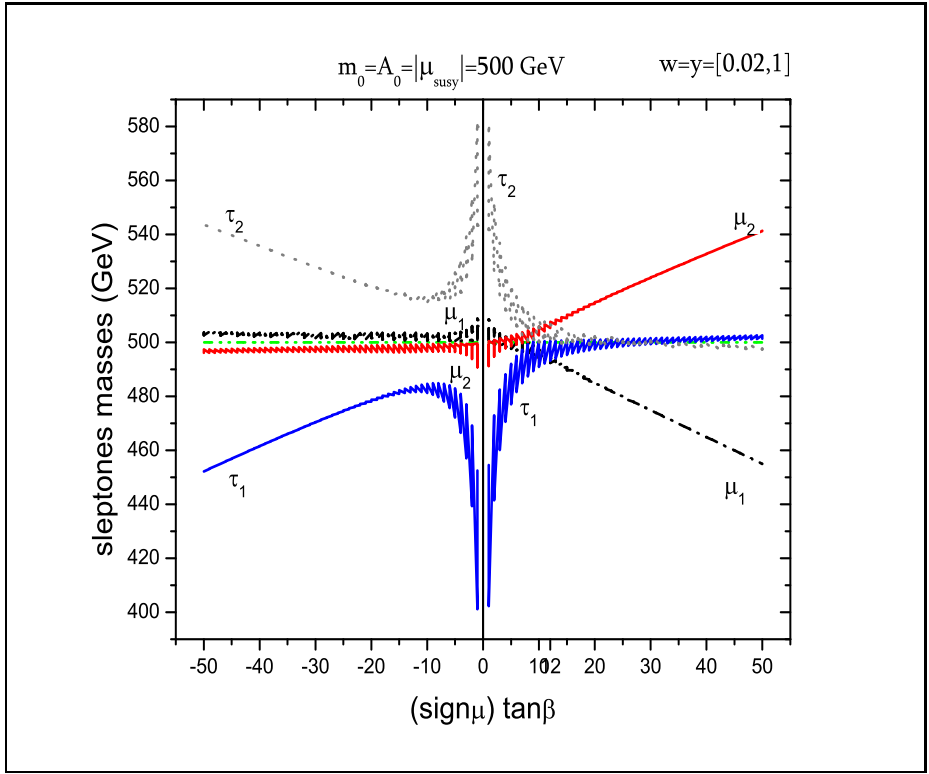}}
\caption[]{\label{fg:BtanB} {\it {\small Slepton masses with respect
to  $\tan \beta$ for $\mu_{susy}<0$, $\mu_{susy}>0$ and with
$w=y=[0.02,1]$, $\tilde{m}_{0}=A_{0}=|\mu_{susy}|=500\,GeV$.}}}
\end{center}
\end{figure}

In figure \ref{fg:BtanB} we verify the behavior of slepton masses
with $\tan\beta$, we run the ansatz parameter through the interval $y=w=[0.02,1]$,
and $\tilde{m}_{0}=500\,GeV$. We found that for $\mu <0$ the smuons
are independent for $\tan\beta > 5 $, while for $\mu >0$ the staus are the ones
independent, but for $\tan \beta > 15$.\\
\begin{figure}[hbt!]
\begin{center}
\framebox {\includegraphics[width=13cm,height=12cm]{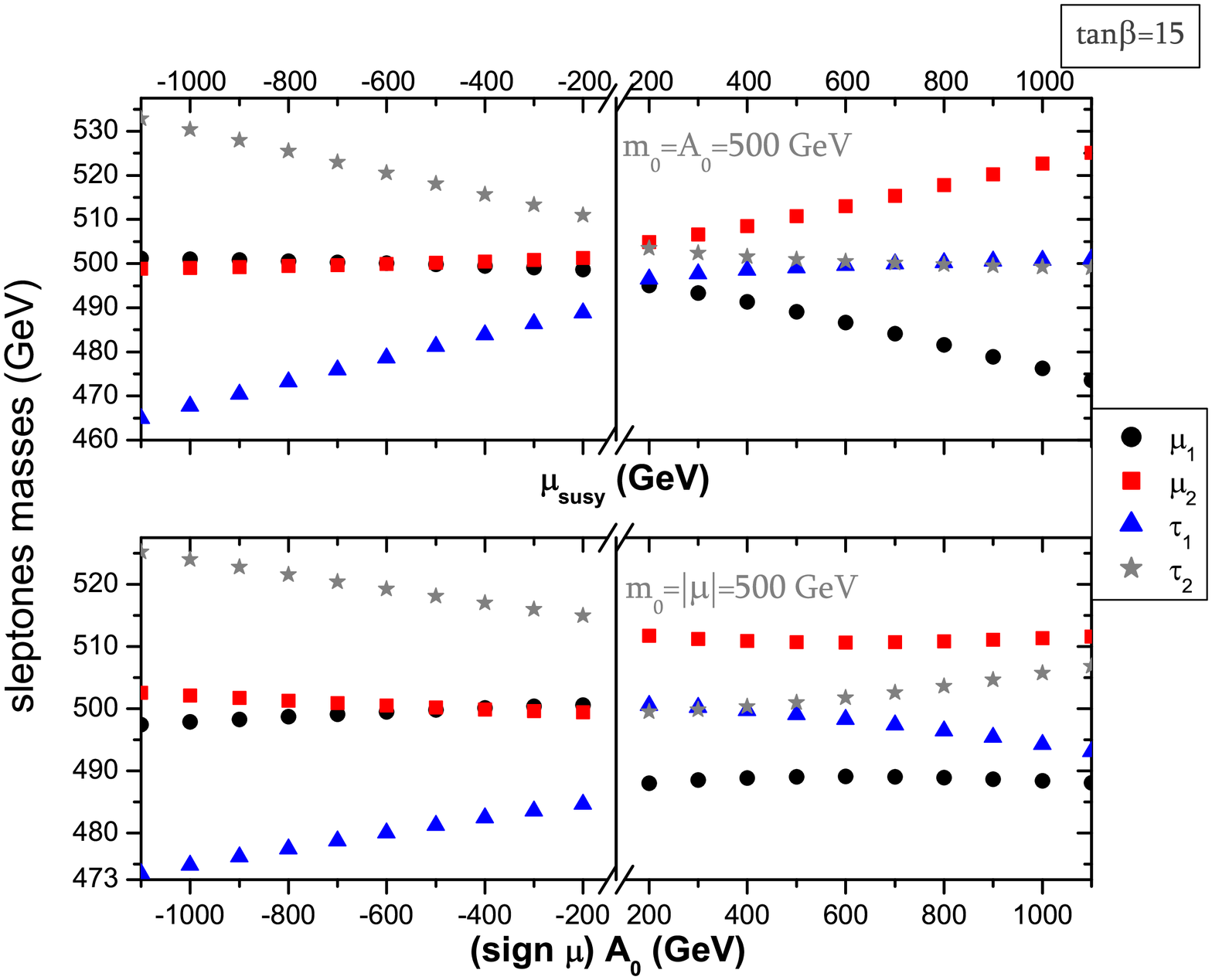}}
\caption[]{\label{fg:BmuA0} {\it {\small Slepton masses dependance
on $\mu_{susy}$, with $\tilde{m}_{0}=A_{0}=500\, GeV$ (up). And
slepton masses dependence on $A_0$ for $\mu_{susy}<0$ and
$\mu_{susy}>0$, with $\tilde{m}_{0}=|\mu_{susy}|=500\, GeV$ (down).
Both with $\tan \beta=15$ and $y=w=1$.}}}
\end{center}
\end{figure}

Although we have considered the all SSusyB parameters equal to the SUSY
breaking mass scale $\tilde{m}_{0}=|\mu|=A_{0}$, this is not
necessary true. We explore independently the possible values for the
Higgsinos mass parameter $\mu$ from the soft mass term as is shown
in top of figure \ref{fg:BmuA0}. In the same sense we explore
independently the $trilineal-A$ coupling, the results are shown in
the bottom of the same figure, \ref{fg:BmuA0}. In both cases we set
the soft mass term as $\tilde{m}_{0}=500\, GeV$. We observed again
the difference in the mass hierarchy between smuons and staus
depending on the $\mu$ sign. In the trilinear coupling dependency, we
observe that the non-degeneration increases for $A_{0}> \tilde{m}_{0}$.\\

\section{Conclusions}

We have study the possible non-degeneration for the sleptones
massses, using an Ansatz for the slepton mass matrix. Specifically,
consider the mixing to occur between the second and third families,
and assume that this mixing comes solely from left-right terms. We
encounter the parameter space dependency of the masses, including
both the MSSM parameters and the proposed model parameters. This
non-degeneracy could be measured in the cases where it is about
$5\%$ of the SUSY Soft-Breaking mass scale $\tilde{m}_{0}$, this
percentage is suggested by considering the
experimental uncertainty.\\

We observed that the strongest dependence comes from the MSSM
parameter space. While, as we expected, the parameters of the ansatz
act only to accomplish for some non-zero terms.\\

A dependence on $\mu$ sign is strongly manifested. The mass
hierarchy changes whether $\mu$ is positive or negative, this lead
us to the conclusion that if the hierarchy mass spectrum most
expected, {\it i.e.}
$m_{\tilde{\tau}_{1}}<m_{\tilde{\mu}_{1}}<m_{\tilde{\mu}_{2}}<m_{\tilde{\tau}_{2}}$
then $\mu$ must be negative.
 Also we observed that for each case,

\begin{itemize}

\item For \emph{$\mu < 0$}, we obtain non-degeneration on staus, with a
difference between them of 10\% or more, for $\tan\beta \sim> 30$
and $|\mu|/\tilde{m}_{0}\sim > 1.6$. And we have practically smuons
degeneration. In this case, considering $A_{0}/\tilde{m}_{0}
> 2 $, generates a difference in stau masses
of $ \sim 10\%$ of $\tilde{m}_{0}$, with $\tan\beta =15$ while for
the smuons we reach only $1\%$. For the ansatz parameters we also
have an increase in  mass difference up to $2\%$ for $y=w=1$

\item For \emph{$\mu > 0$}, the non-degeneration is obtained for the smuons, and the difference between
$\tilde{\mu}_{1}$ and $\tilde{\mu}_{2}$, could be larger than 10\%
for $\tan\beta \sim> 30$ and $|\mu|/\tilde{m}_{0}\sim > 2$, while we
obtain approximately stau degeneration, where only for
$A_{0}/\tilde{m}_{0}> 2 $, we reach a difference of $ > 3\%$ of the $\tilde{m}_{0}$.\\
Analyzing the ansatz parameters, we obtain an increased mass
difference for $y=w=1$ getting up to $2\%$, with the strongest
dependency being on the $w$ parameter.\\
\end{itemize}

For $\tan\beta$ we conclude that if a degenerated masses are
measured then $\tan\beta$ value should be at around $10$, while in
the other case, no-degeneration is manifested either at small
$tan\beta$, (less than $\sim 5$) or for large value.\\

The mass difference found here could be tasted possible at LHC, with
some difficulties, but certainly at the ILC.\\

\end{document}